\documentclass[twocolumn]{aastex631}

\received{XXX}
\revised{YYY}
\accepted{ZZZ}

\shorttitle{Detection of the secondary in OJ287}
\shortauthors{Valtonen et al.}

\graphicspath{{./}{}}

\begin{document}

\title{Evidence of jet activity from the secondary black hole in the OJ~287 binary system}

\author[0000-0001-8580-8874]{Mauri J. Valtonen}
\affiliation{FINCA, University of Turku, FI-20014 Turku, Finland}
\affiliation{Tuorla Observatory, Department of Physics and Astronomy, University of Turku, FI-20014 Turku, Finland}

\author[0000-0003-3609-382X]{Staszek Zola}
\affiliation{Astronomical Observatory, Jagiellonian University, ul. Orla 171, 30-244 Krakow, Poland; szola@oa.uj.edu.pl}

\author[0000-0002-9331-4388]{Alok C. Gupta}
\affiliation{Aryabhatta Research Institute of Observational Sciences (ARIES), Manora Peak, Nainital -- 263001, India}
\affiliation{Key Laboratory for Research in Galaxies and Cosmology, Shanghai Astronomical Observatory, Chinese Academy of Sciences, 80 Nandan Road, Shanghai 200030, People's Republic of China}

\author{Shubham Kishore}
\affiliation{Aryabhatta Research Institute of Observational Sciences (ARIES), Manora Peak, Nainital -- 263001, India}

\author{Achamveedu Gopakumar}
\affiliation{Department of Astronomy and Astrophysics, Tata Institute of Fundamental Research, Mumbai 400005, India}

\author[0000-0001-6158-1708]{Svetlana G. Jorstad}
\affiliation{Institute for Astrophysical Research, Boston University, 725 Commonwealth Avenue, Boston, MA 02215, USA}

\author[0000-0002-1029-3746]{Paul J. Wiita}
\affiliation{Department of Physics, The College of New Jersey, 2000 Pennington Rd., Ewing, NJ 08628-0718, USA}

\author[0000-0002-4455-6946]{Minfeng Gu}
\affiliation{Key Laboratory for Research in Galaxies and Cosmology, Shanghai Astronomical Observatory, Chinese Academy of Sciences, 80 Nandan Road, Shanghai 200030, People's Republic of China}

\author{Kari Nilsson}
\affiliation{FINCA, University of Turku, FI-20014 Turku, Finland}

\author[0000-0001-7396-3332]{Alan P. Marscher}
\affiliation{Institute for Astrophysical Research, Boston University, 725 Commonwealth Avenue, Boston, MA 02215, USA}

\author[0000-0002-8366-3373]{Zhongli Zhang}
\affiliation{Shanghai Astronomical Observatory, Chinese Academy of Sciences, Shanghai 200030, People's Republic of China}
\affiliation{Key Laboratory of Radio Astronomy and Technology, Chinese Academy of Sciences, A20 Datun Road, Chaoyang District, Beijing 100101, People's Republic of China}

\author{Rene Hudec}
\affiliation{Faculty of Electrical Engineering, Czech Technical University, 166 36 Prague, Czech Republic; rene.hudec@gmail.com}
\affiliation{Astronomical Institute (ASU CAS), 251 65 Ond\v{r}ejov, Czech Republic}

\author{Katsura Matsumoto}
\affiliation{Astronomical Institute, Osaka Kyoiku University, 4-698 Asahigaoka, Kashiwara, Osaka 582-8582, Japan; katsura@cc.osaka-kyoiku.ac.jp}

\author{Marek Drozdz}
\affiliation{Mt. Suhora Astronomical Observatory,  University of the National Education Commission,
ul.Podchorazych 2, 30-084 Krakow, Poland}

\author{Waldemar Ogloza}
\affiliation{Mt. Suhora Astronomical Observatory,  University of the National Education Commission,
ul.Podchorazych 2, 30-084 Krakow, Poland}

\author{Andrei V. Berdyugin}
\affiliation{Tuorla Observatory, Department of Physics and Astronomy, University of Turku, FI-20014 Turku, Finland}

\author{ Daniel E. Reichart}
\affiliation{Department of Physics and Astronomy, University
 of North Carolina at Chapel Hill, Chapel Hill, NC 27599, USA; dan.reichart@gmail.com }

\author{Markus Mugrauer}
\affiliation{Astrophysical Institute and University Observatory, Schillergässchen 2, D-07745 Jena, Germany; markus@astro.uni-jena.de}

\author{Lankeswar Dey}
\affiliation{Department of Physics and Astronomy, West Virginia University, PO Box 135, Willey Street, Morgantown, WV 26506
USA}

\author{Tapio Pursimo}
\affiliation{ Nordic Optical Telescope, Apartado 474, E-38700 Santa Cruz de La Palma, Spain; tpursimo@not.iac.es}

\author{Harry J. Lehto}
\affiliation{Tuorla Observatory, Department of Physics and Astronomy, University of Turku, FI-20014 Turku, Finland}

\author[0000-0002-0712-2479]{Stefano~Ciprini}
\email{stefano.ciprini@ssdc.asi.it}
\affiliation{Istituto Nazionale di Fisica Nucleare, Sezione di Roma
``Tor Vergata", I-00133 Roma, Italy}
\affiliation{Space Science Data Center - Agenzia Spaziale Italiana, Via
del Politecnico, snc, I-00133, Roma, Italy}

\author{T. Nakaoka}
\affiliation{Hiroshima Astrophysical Science Center, Hiroshima University, 1-3-1 Kagamiyama, Higashi-Hiroshima, Hiroshima 739-8526, Japan}

\author[0000-0002-7375-7405]{M. Uemura}
\affiliation{Hiroshima Astrophysical Science Center, Hiroshima University, 1-3-1 Kagamiyama, Higashi-Hiroshima, Hiroshima 739-8526, Japan}

\author[0000-0002-0643-7946]{Ryo Imazawa}
\affiliation{Department of Physics, Graduate School of Advanced Science and Engineering, Hiroshima University, 1-3-1 Kagamiyama, Higashi-Hiroshima, Hiroshima 739-8526, Japan}

\author{Michal~Zejmo}
\affiliation{ Kepler Institute of Astronomy, University of Zielona Gora, Lubuska 2, 65-265 Zielona Gora, Poland; michalzejmo@gmail.com}

\author{Vladimir~V.~Kouprianov}
\affiliation{ Department of Physics and Astronomy, University of North Carolina at Chapel Hill, Chapel Hill, NC 27599, USA; v.kouprianov@gmail.com }

\author[0009-0007-1284-7240]{James W. Davidson, Jr.}
\affiliation{Department of Astronomy, University of Virginia, 530 McCormick Rd., 
Charlottesville, VA, 22904, USA; jimmy@virginia.edu}
 
\author{Alberto Sadun}
\affiliation{ Department of Physics, University of Colorado, Denver, CO 80217, USA; alberto.sadun@ucdenver.edu}

\author{Jan \v{S}trobl }
\affiliation{Astronomical Institute (ASU CAS), 251 65 Ond\v{r}ejov, Czech Republic}

\author[0000-0001-6314-0690]{Z. R. Weaver}
\affiliation{Institute for Astrophysical Research, Boston University, 725 Commonwealth Avenue, Boston, MA 02215, USA}

\author{Martin Jel\'{\i}nek}
\affiliation{Astronomical Institute (ASU CAS), 251 65 Ond\v{r}ejov, Czech Republic}

\begin{abstract}
\noindent

We report the study of a huge optical intraday flare on November 12, 2021, at 2 am UT, in the blazar OJ~287. In the binary black hole model it is associated with an impact of the secondary black hole on the accretion disk of the primary. Our multifrequency observing campaign was set up to search for such a signature of the impact, based on a prediction made eight years earlier. The first I-band results of the flare have already been reported by \cite{2024ApJ...960...11K}. Here we combine these data with our monitoring in the R-band. There is a big change in the R-I spectral index by $1.0\pm0.1$ between the normal background and the flare, suggesting a new component of radiation. The polarization variation during the rise of the flare suggests the same. The limits on the source size place it most reasonably in the jet of the secondary black hole. We then ask why we have not seen this phenomenon before. We show that OJ~287 was never before observed with sufficient sensitivity on the night when the flare should have happened according to the binary model. We also study the probability that this flare is just an oversized example of intraday variability, using the Krakow-dataset of intense monitoring between 2015 and 2023. We find that the occurrence of a flare of this size and rapidity is unlikely. In the Appendix, we give the full orbit-linked historical light curve of OJ~287 as well as the dense monitoring sample of Krakow.
\end{abstract}

\section{Introduction} \label{sec:introduction}
\noindent
OJ~287 is a highly variable BL Lacertae type quasar at a rather low redshift of 0.306 \citep{1985PASP...97.1158S}. It is easily observable even with small telescopes. Because it lies close to the ecliptic, it has been accidentally photographed since 1887 during searches of minor planets and other objects near the ecliptic plane. This has produced a vast amount of data: several hundred photometric measurements and interesting upper limits prior to 1970. The magnitude data displays an easily discernible 12 year cycle, modified by another 55 yr cycle. These come out prominently in quantitative analysis, but are easily seen by just looking at the light curve (Fig. 1). In addition, the dense network of upper limits puts severe restrictions on the light curve in the parts where the photometric data are sparse \citep[e.g.,][and references therein]{2021Galax..10....1V,2023MNRAS.521.6143V}. 
\begin{figure}
    \centering
    \includegraphics[angle = 0,width=0.47\textwidth]{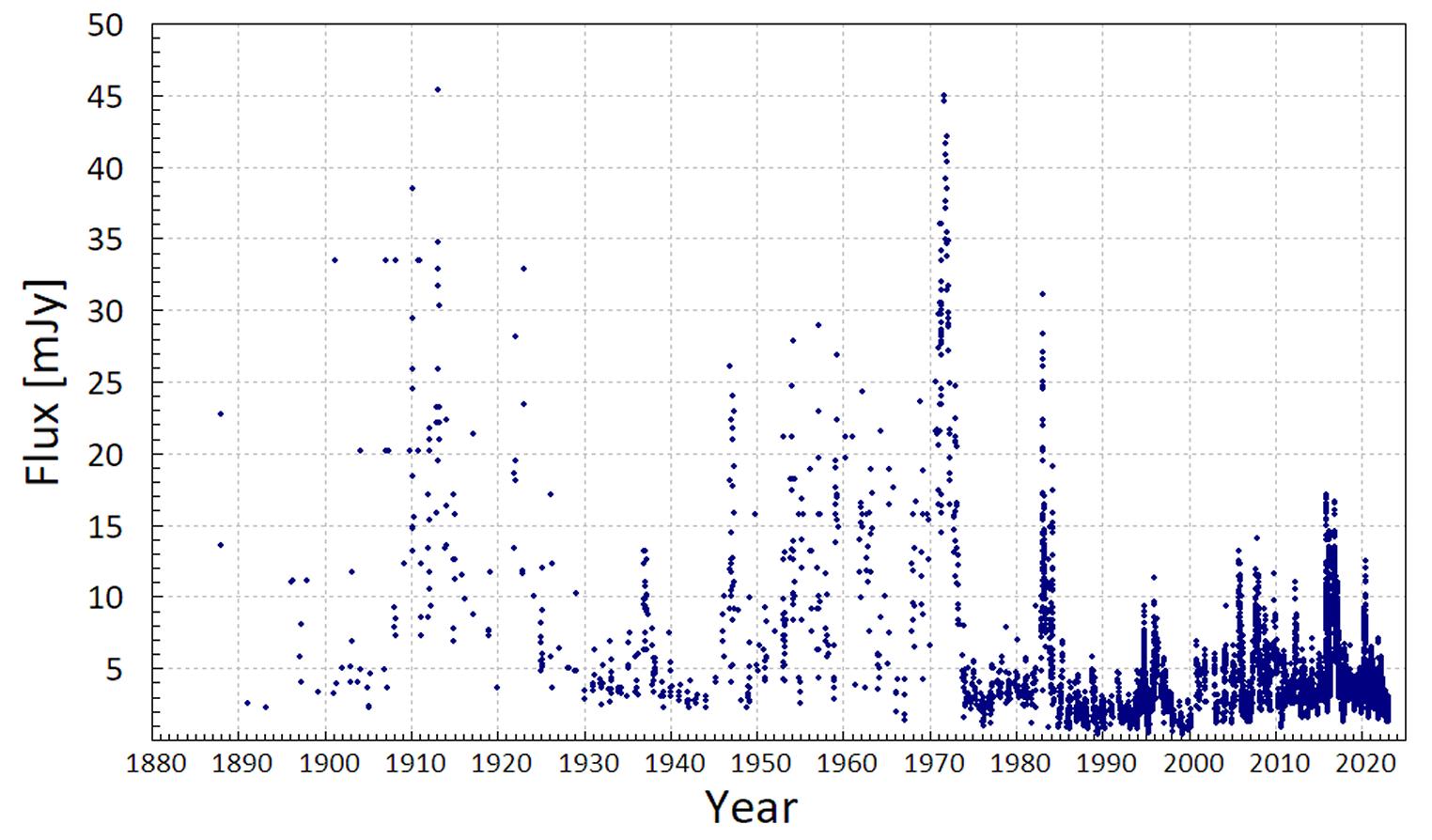}
    \caption{V-band historical light curve of OJ~287}.
\label{fig:illu}
\end{figure}

The understanding of the two light curve cycles as well as the exact times of the large flares is paramount to any theoretical model. The large flares represent typically a two magnitude rise in brightness. In addition to the flares seen in the photometric data, there are only a very limited number of flares that can exist without violating the upper limit network. A mathematical sequence, called Keplerian sequence, similar in nature to the Balmer formula for the hydrogen spectral lines, gives all the observed flares seen in OJ~287, and does not contradict the upper limits of the photometry of OJ~287 \citep{2023MNRAS.521.6143V}. There are currently 26 members in this sequence starting from 1886 up-to-date. The sequence is based on a simple analytical model of binary motion.

The sequence provides a way of predicting future flares with the accuracy of about one year. All it requires is the assumption that the system consists of an unequal mass binary black hole (BH) pair and a gas disk. The signals arise from the plunging of the secondary BH through the gas disk.

An even simpler method is to take a stretch of the old light curve of OJ~287 and to slide it forward on the time axis to future time. It does not necessarily need any astrophysical theory to back it up. It is a purely phenomenological method, and all it requires is that there is some sort of repeatability in the system. This method has been used to justify observing campaigns during specified periods of time. The past experience is that the sliding-on-the-time-axis method is useful with about one year accuracy, i.e. it can be used to justify the starting of observing campaigns over a specific observing season independent of any astrophysical theory.

Then there are highly accurate models which predict the times of the flares with the accuracy of up to four hours \citep{2020ApJ...894L...1L}. The following values for OJ~287's BH binary system are found:
primary mass $m_1 = 18.35\pm0.05 \times 10^9 M_{\odot}$, secondary mass $m_2 = 150\pm10 \times 10^6 M_{\odot}$, primary Kerr parameter $\chi_1 = 0.38\pm0.05$, orbital eccentricity $e = 0.657\pm0.003$, and orbital period (redshifted) $P = 12.06\pm0.01$ years.

The Keplerian sequence is just an example of approximate solutions to the timings of the flares. The full solution was calculated by  using a binary black hole (BH) model in General Relativity up to Post-Newtonian order 4.5, spin-orbit interaction, a standard accretion disk of two parameters, and accurate calculations of disk bending. Also, the feedback of the disk potential onto the binary orbit was included in a self-consistent way, even though it was found to be insignificant \citep{1988ApJ...325..628S,1997ApJ...484..180S}. The direction of the observer relative to the disk was also taken into account \citep{1998ApJ...507..131I}. 

The total number of parameters is eight. The method solves the parameters by a convergent method similar to the Newton-Raphson method. The method requires the exact timings of nine flares, and solves the correct times for another eight flares which have been sufficiently observed. In addition, it satisfies strict observational constraints in many cases where a flare was not seen, but the region of the light curve is densely covered by upper limits. Even though the convergent method is efficient, it still requires millions of orbit solutions to find out the range of uncertainties of the parameters, in addition to their most likely values.

All the free parameters are related to the astrophysical model. The theory of gravity and its numerical treatment are given, and contain no parameters. In this way the method differs from \cite{2023MNRAS.526.2754} who treat the theory of gravity by a parametric method, and determine the parameters by fitting to OJ~287 observations. Then it remains unclear how the method differs  from the standard gravitational theory.

In principle the solution could fall on a false minimum rather than on the global minimum, as suggested by \cite{2023MNRAS.526.2754}. However, if we look at the history of development of this model from 1995 to 2018, we see that the primary properties of the model (binary masses, orbital eccentricity, etc.) have not changed when greater astrophysical details have been included and the number of free parameters has increased from four to eight.  The purely mathematical Keplerian sequence, with no astrophysical details beyond Newton's law of gravity and Einstein's explanation of the first order orbit precession, leads to the same orbit solution. The simplified model suggested by \cite{2023MNRAS.526.2754}, even though lacking essential astrophysical details such as disk bending, direct calculation of disk potential by full N-body simulation, spin-orbit interaction, details of radiation processes, etc., leads to essentially the same solution of the main properties of the system as in the full solution. 

The model was completed in 1995 and was presented in several papers thereafter. We call it the \textit{standard model} in the following. It combines three accurate codes: the code of orbit calculation \citep{2020gfbd.book.....M}, the code of disk potential calculation \citep{1976JCoPh..21..400M} and the code of evolution of an expanding gas cloud, developed by Harry Lehto \citep{1996ApJ...460..207L}. These codes were combined into a single code by Björn Sundelius for the application to the OJ287 problem \citep{1997ApJ...484..180S}. This code also had a feature of either keeping the self-interaction inside the disk or removing it when it was not necessary, in order to speed up the calculations. We should note that very similar codes are widely used in solving solar system problems where their functioning can be directly verified by observations \citep{https://doi.org/10.1143/PTPS.195.48}. 

With the later addition of disk-sidedness (that is, from which side we view the disk), spin-orbit interaction, disk bending, and the General Relativistic tail-terms, we may say that the full solution of the OJ287 problem is currently comparable in accuracy to studies in the solar system. For example, we are able to predict the times of new flares in OJ~287 with the same relative accuracy as the next apparition of Halley's comet in the inner solar system. 

The omission of the spin-orbit interaction and the disk bending both lead to errors in excess of 0.5 yr in the times of flares \citep{2007ApJ...659.1074V,2011ApJ...742...22V}. Thus such models are of little interest today, besides proving that the full mathematical solution of the OJ~287 problem is unique.

The full solution was tested in 2019 when the observed flare came within four hours of the predicted time \citep{2020ApJ...894L...1L}, and again in 2022, when the observations put strict limits on the timing even though the flare itself was not observable \citep{2018ApJ...866...11D,2023MNRAS.521.6143V}.

The graphical sliding-on-the-time-axis approach was most recently used for the 2021-2022 multifrequency campaign. The usefulness of this method is that it does not use any specific model, but gives an idea when to carry out observations on general grounds. These ideas were communicated in one paper \citep{2021Galax..10....1V} and in various pre-publication notes (dated June, September and November 2022). In the final publication it was shown that one must use multicolor data for the sliding to get useful results \citep{2023MNRAS.521.6143V}.  The prediction in the standard model was given in \cite{2018ApJ...866...11D}. Using the R-band data alone produces a flare date which was far too late with respect to the accurate model, while the B-band data gives excellent agreement. 

The light curve comparison between years 2005 and 2022 also produced an important piece of additional information which we did not have before. The disk crossing times require the knowledge of the disk level, i.e. how much the disk is bent above or below its mean level at the time and at the position of the disk crossing. Also the astrophysical delay from the disk crossing to the observed flare has to be calculated. In principle there could be errors in both quantities which accidentally cancel each without affecting the observed flare time. \cite{2023MNRAS.521.6143V} showed that the astrophysical time delay can be directly measured in the 2005 disk impact, and the measurement agrees with the standard model within errors. 

The disk level calculation uses only low-level General Relativistic corrections to the Newtonian theory. The corrections are quite standard; higher level corrections are required for the binary motion but not for the disk motion since we are not looking at the central parts of the disk. Comparing the simulations of the disk level between 2005 and 2022 disk impacts, it was found that the uncertainty is about 5 AU \citep{2023Galax..11...82V}. In terms of travel time, this corresponds to about 12 hours at these impact distances. For disk impacts at closer distance from the central BH (i.e., closer than 10,000 AU) the mean level is zero, and the standard error of the mean is about 4 AU \citep{2007ApJ...659.1074V}. The orbital speed at the pericenter is also higher than in the apocenter part of the orbit, and therefore the disk level uncertainty is below 4 hours in the timing the pericenter flares. This is better than what can be determined from observations which means that the disk level uncertainties play no role in the standard model.

However, it is important to differentiate between the full mathematical solution which we call the \textit{standard model}, and the graphical method. The reason why the graphical method works only with the B-band data is that the effects of the disk impact show primarily in the blue color while the R-band data is dominated by the primary jet. The two types of activity take place in different regions of the system, and cannot be simply connected \citep{2023Galax..11...82V}.

The observation of the main flare in 2022 was known to be impossible by ground-based optical telescopes. Radio observations were carried out, and they excluded the possibility that the observed flare sequence arises from activity in the primary jet. If it did, we would have seen a radio flare in the summer of 2022, but there was no evidence of it.

However, there is another way to get essentially the same information from the optical light curve. In the model of \cite{2013ApJ...764....5P} it has been argued that the secondary BH becomes active at certain phases of the binary orbit, and these active stages can be used as orbit markers. If seen at the predicted times, this activity of the secondary component also confirms the orbit model. 

The standard flares are interpreted as a result of the impacts of the secondary on the accretion disk. This causes thermal radiation from bubbles of gas which are pulled out of the disk. In contrast, the flares from the secondary itself would be associated with the jet of the secondary BH. This emission would appear on top of the normal emission from the primary jet during brief periods of time when the Roche lobe of the secondary is flooded by the disk gas. Even though the secondary is 122 times smaller in mass than the primary, during these special episodes the emission of the secondary jet can overpower the emission of the primary jet and show up as large amplitude intraday variability (IDV). 

This paper reports very dense photometry of OJ~287 during the period when the exact orbit solution makes us expect the jet emission from the secondary BH. It is complemented by polarimetric observations, which are less dense but can also provide useful information. We then compare our data with \cite{2024ApJ...960...11K} who confirmed the appearance of the flare which we had already tentatively reported \citep{2023AAS...24232203V}. We then discuss our findings with respect to the theoretical model calculated by \cite{2013ApJ...764....5P}.  Finally we use the model orbit file from 1887 onward and and compare it with the available photometry of OJ~287, and ask whether the secondary BH flares have been seen previously at corresponding times. The significance of our findings are discussed in the conclusions.

\section{Data Collection and Reduction in the R-band}
\noindent
For the flares which we usually discuss in OJ~287, the full width at half-maximum is from a few days to weeks \citep{2008Natur.452..851V}. The question we pose here is whether even faster IDV flares exist. Until this campaign, none had been reported in OJ~287 in the same brightness category as the ordinary big flares ($\geq 4$ mJy in the R-band \citep{1996ApJ...460..207L}). But it does not mean that they do not exist. The shorter the flare time scale, the harder it is to catch them. And since we have a 12 year periodicity, it is easy to see how a one-night event per 12 yr could be missed. It becomes crucial that there is already a reliable model which tells us when to organize a campaign.

The first attempt for such a campaign took place in 2013, with the inspiration of \cite{2013ApJ...764....5P}. The Krakow monitoring program was active, but not dense enough. The next chance came at the end of 2021, and at this time we were better prepared, as we will now describe.

\begin{figure}
    \centering
    \includegraphics[angle = 270,width=0.47\textwidth]{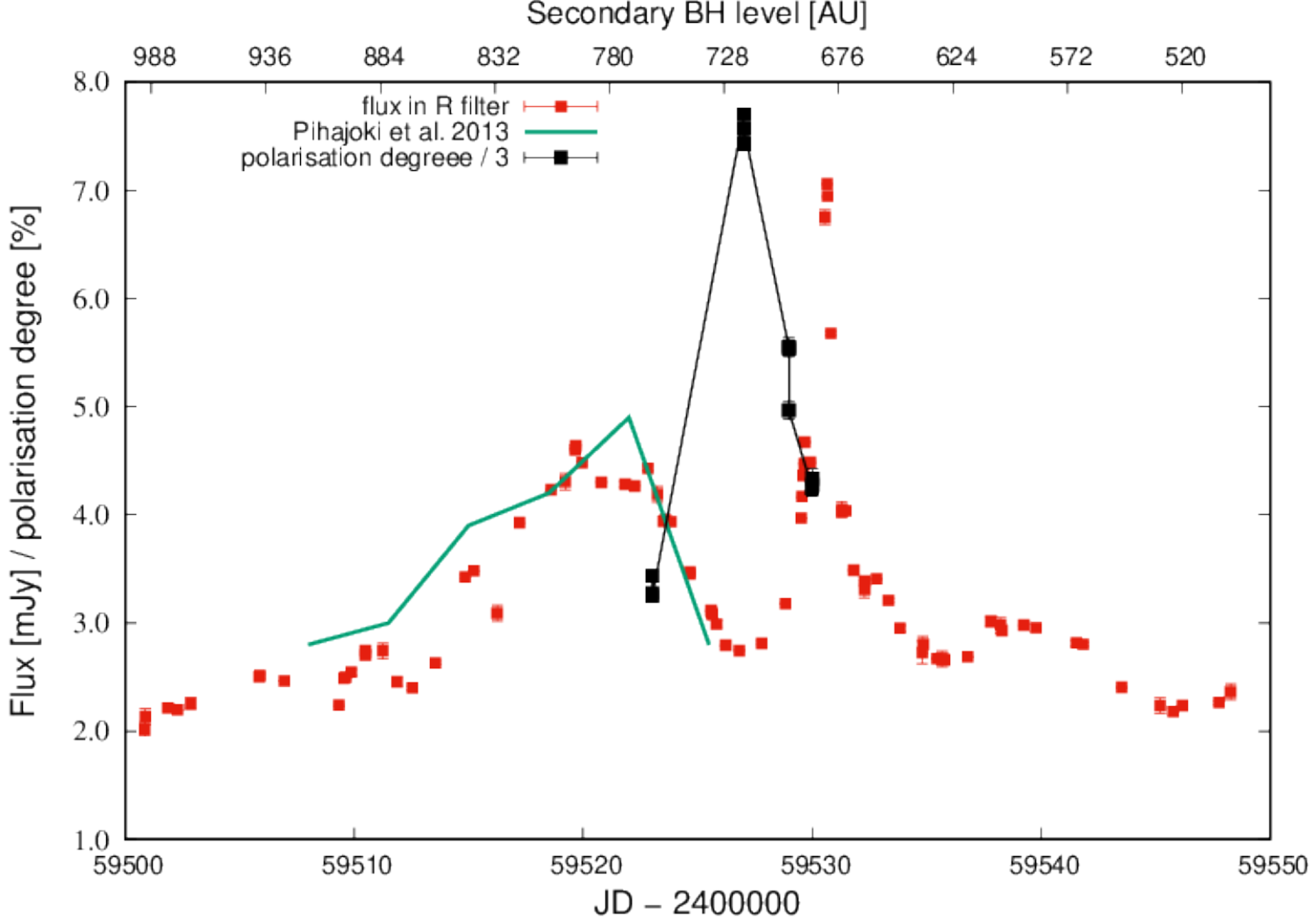}
    \caption{The R-band data (red squares), degree of polarization measurements (black squares) and a theoretical line from Pihajoki et al. (2013). The time in Julian Days is given below, and the corresponding positions of the secondary BH above the disk are labelled on the top.
}
    \label{fig:magVsPDPA}
\end{figure}

Photometric monitoring of OJ~287 within the Krakow Quasar Monitoring Program
started in 2006. Initially we performed observations at two Polish sites:
Mt. Suhora Observatory of the Pedagogical University in Krakow and the
Astronomical Observatory of the Jagiellonian University. Observations taken at the two nearby sites suffered gaps in data, mainly due to bad weather. Therefore, in 2013 the monitoring program started to use the prompt5 telescope, located in Chile and
controlled by the Skynet Robotic Telescope Network \citep{2021RMxAC..53..206Z}. Since 2015, we used seven optical telescopes within the Skynet Network. Due to their location on four continents and the telescope redundancy, we could gather data daily. Additional observations are being provided routinely by the Osaka Astronomical Observatory and the University Observatory Jena. During multisite campaigns, several other sites, located all over the globe, contributed photometric data which have previously been reported (\citet{2023MNRAS.521.6143V} and references therein). 
Images were reduced in a standard way: calibration of raw images was done using
the {\it IRAF} package, while differential magnitudes were extracted with the {\it CMunipack} program, which is an interface to the {\it DAOPHOT} code. The data taken
by the Skynet telescopes have been calibrated by the network pipeline.
We used stars \#4  and \#10 \citep{1996A&AS..116..403F} as comparison and check stars, respectively. 
The photometric data reported and analyzed in this paper cover the period
between Oct 21 and Dec 1, 2021, when a huge flare in the optical band has occurred.
The data taken in the wide band R filter within this period consist of
562 single points binned into 42 mean ones with a 1 day bin. We adopted the brightness of the comparison star in the R filter as 13.74 mag \citep{1996A&AS..116..403F} and converted magnitudes into flux in mJy units. These data are shown in Figure \ref{fig:magVsPDPA} by red squares.

Altogether, until July 24, 2023 we gathered 63480 individual points, which were binned 
with 12 hours bins, resulting in 3388 mean points. We make the data available 
in Table \ref{TabAllRData}. Due to its length, we show here only the first
10 entries, and the entire Table is available in the electronic form only. 
Individual data are also available on request.  

\begin{table}
\begin{center}
\caption{OJ~287 mean points gathered in the R filter during the period between Sep 30, 2006 and Jun 24, 2023. For the brightness of the comparison star, see \cite{1996A&AS..116..403F}. Individual measurements were binned with 12 hours bins. The entire dataset is available in electronic form only.}   
\begin{tabular}{lcccr}
\hline
         JD$_{hel}$   &    OJ287-comp [mag] & $\sigma$ & NPTS \\
\hline         
     2454008.598390   &     1.557    &     0.003  &  40 \\
     2454039.645830   &     1.204    &     0.005  &  21 \\ 
     2454047.617653   &     1.821    &     0.005  &  37 \\
     2454073.528083   &     1.357    &     0.007  &   9 \\
     2454075.480830   &     1.368    &     0.030  &   3 \\
     2454075.647513   &     1.375    &     0.004  &  14 \\
     2454081.462389   &     1.611    &     0.004  &  11 \\
     2454081.517999   &     1.622    &     0.003  &  13 \\ 
     2454084.450372   &     1.575    &     0.002  &  16 \\ 
     2454084.534483   &     1.581    &     0.007  &   6 \\
\hline
\end{tabular}
\label{TabAllRData}
\end{center}
\end{table}

\section{Comparison between TESS I-band and our R-band observations}
\begin{figure}
    \centering
    \includegraphics[angle = 0,width=0.47\textwidth]{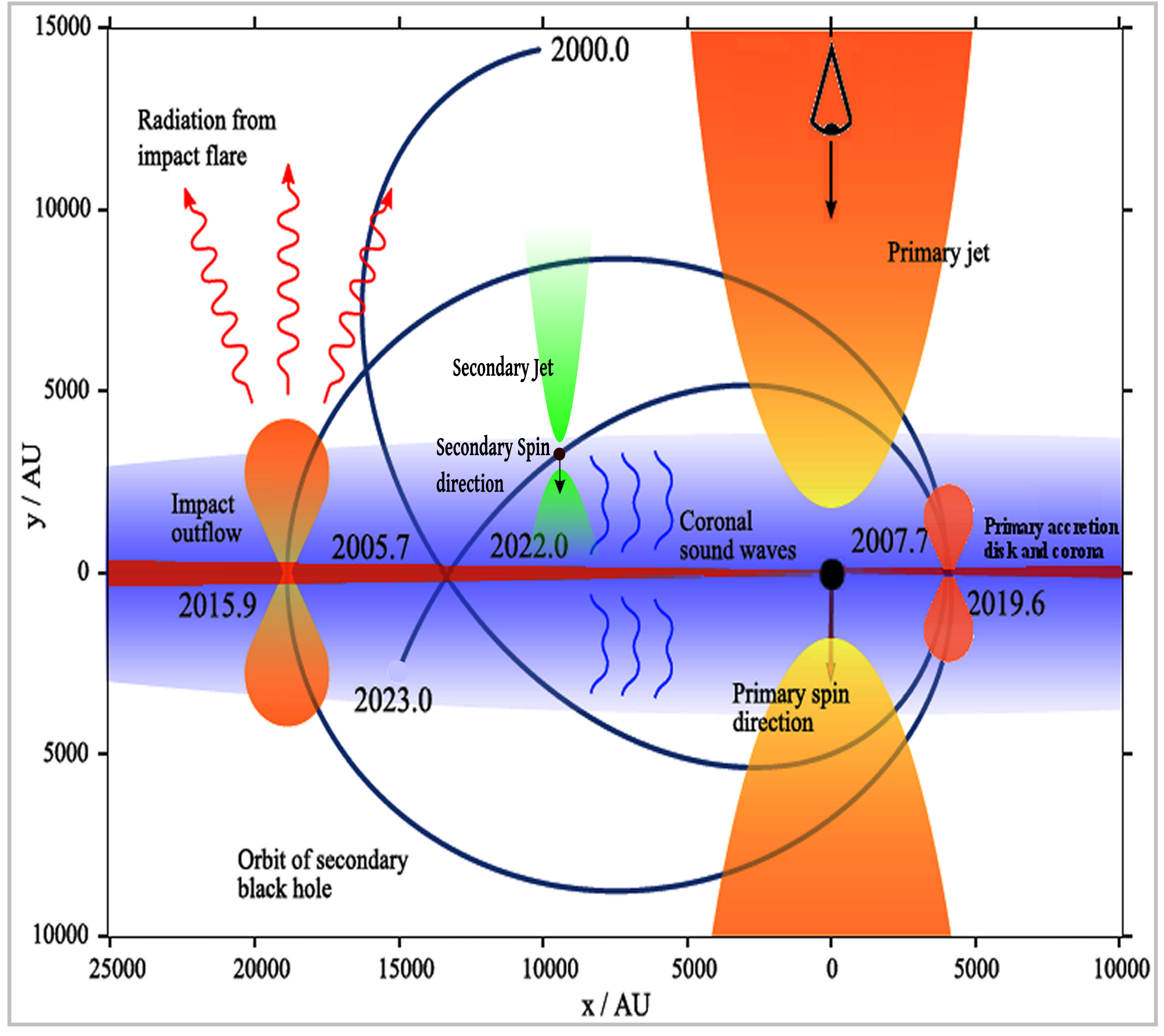}
    \caption{The binary model of OJ~287. The secondary jet pointing at us is thought to be the source of the November 12, 2021, flare. The size of the radiating region is comparable to the expected cross-section of this jet, but far too small for the cross-section of the primary jet.}
    \label{fig:illu1}
\end{figure}

\noindent
At the end of 2021, during the period of 80 days, the Transiting Exoplanet Survey Satellite (TESS) observed OJ~287 almost continuously. This is the period of time when, according to prior estimates, the secondary should have approached the primary disk from our side, and triggered the secondary jet activity \citep{2013ApJ...764....5P,2021Galax..10....1V}. Figure \ref{fig:illu1} illustrates the situation.

The TESS observations are described in \cite{2024ApJ...960...11K}. The TESS filter covers the traditional Johnson I-band but it is wider. Therefore our first task is to connect the TESS magnitudes T with the Johnson I ones. We do it when OJ~287 was at a low level at the beginning of the TESS run at JD2459500. At this time also the Swift telescope observed it as part of the MOMO project \citep{2022MNRAS.513.3165K}. The spectral energy distribution is found to be quite normal, with the spectral index $\alpha = -1.50\pm0.05$. The $R-T$ color is found to be 0.73, also as expected \citep{1988AJ.....95..307I,2002A&A...381..408E,2018A&A...610A..74K}, if the TESS magnitude corresponds to the Johnson I-band magnitude. We assume in the following that this is the case. The TESS conversion formulae  T = log(counts/s)+20.44, flux = $6.064\times 10^{-0.4(T-14)}$ mJy, are used\footnote{\url{https://tess.mit.edu/public/tesstransients/pages/readme.html}}.

\begin{figure}
    \centering
    \includegraphics[angle = 0,width=0.48\textwidth]{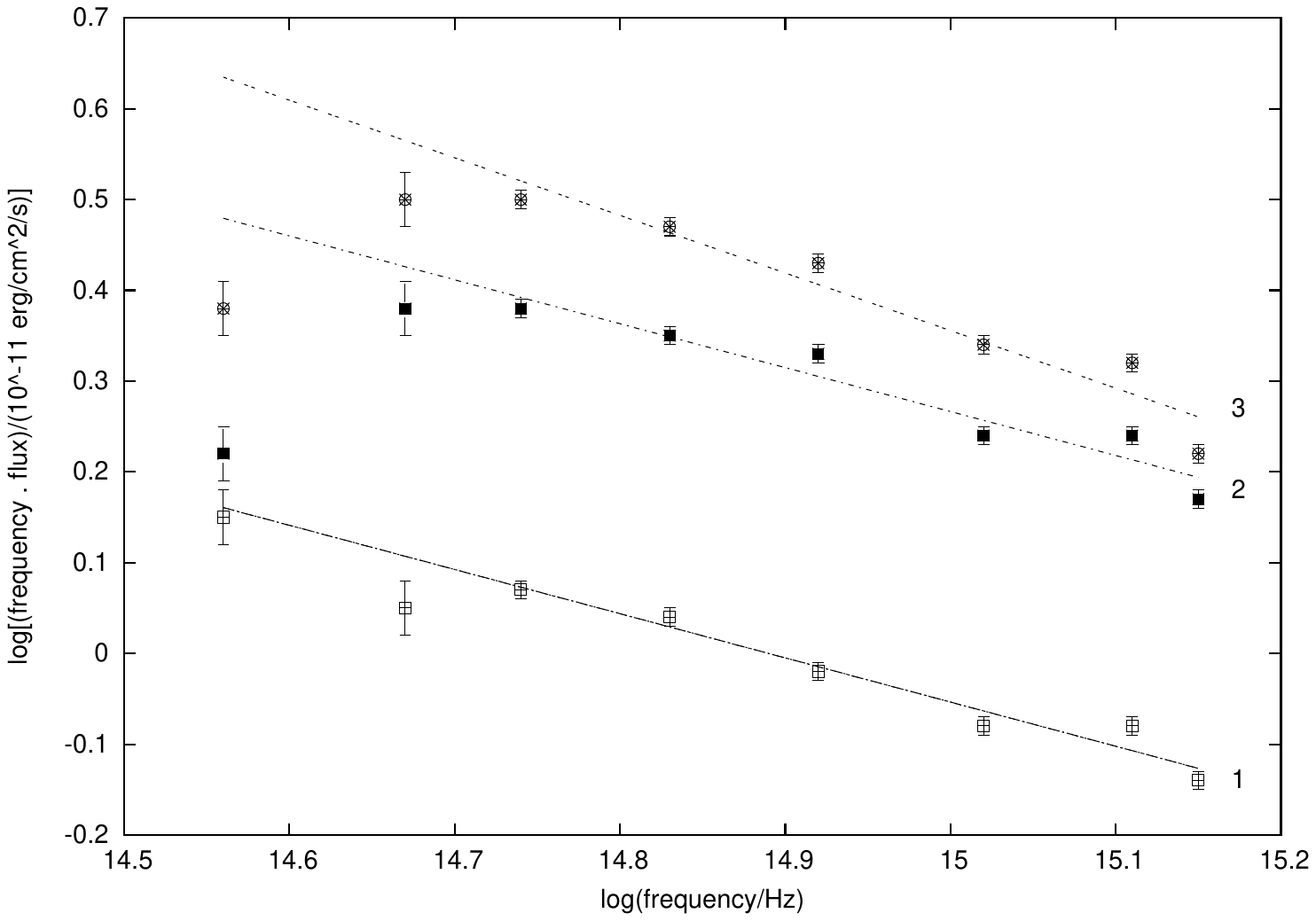}
    \caption{The spectral energy distribution in OJ~287 at the base level (1), the light curve hump (2), and in the early stages of the big flare (3). 
}
    \label{fig:magVsPDPA2}
\end{figure}

The two light curves in the TESS and R bands are similar in general. However, when we calculate the spectral index at different stages of the flare, differences arise. Fig.\ref{fig:magVsPDPA2} shows that during the flares (epochs 2 and 3) the spectral energy distribution peaks around the R-band frequency, in contrast to the base level power-law spectrum where no such peak is seen (epoch 1). At the big flare the spectral index at lower frequencies, between R and I-bands, is $\alpha=-0.49\pm0.04$, while in the background it is $\alpha = -1.50\pm0.05$. Assuming that the base level stayed constant during the 9 day period between the beginning and the end of the flare period, the spectral index of the flare itself is $\alpha\sim0$.

The spectral index differences show up also in colors. At the peak of the flare R-T=0.165$\pm$0.008, while in the hump between JD2459514 and JD2459524 the spectrum flattens: at JD2459519 we find R-T=0.105$\pm$0.03. The last Swift observation in this series was at JD2459530, only 5 hours before the full peak of radiation, when we obtain R-T=0.11$\pm$0.06. The uncertainty in the R-band magnitude is large because the flux was rising fast and the nearest R-band measurement was done 2.5 hours later.

\cite{2024ApJ...960...11K} estimate that the size of the emitting region is $280\pm130$ AU. Comparing this with the Schwarzschild radii of the two black holes in the model, 360 AU and 3 AU, respectively, we see that the emitting region must be located in the jet of the secondary black hole, as jets widen considerably beyond the size of the black hole of their origin \citep{2022ApJ...940...65O,2022ApJ...924..122G,2023Natur.616..686L}. 

Another argument in support of the second component in the radiation during the flare is the behavior of the degree of polarization: while normally the degree of polarization increases with the rising flux \citep{2023ApJ...957L..11G}, in this instance it behaves quite the opposite way (Fig.\ref{fig:magVsPDPA}). This can be understood as superimposing two sources with different polarization properties in the same beam of light. Unfortunately, the polarization measurements did not cover the full radiation peak.

\section{Probability of IDV}
\noindent
Even though the November 12, 2021, flare was the largest ever seen in the intraday (IDV) time scale, the question remains: what is the likelihood that it simply represents the tail end in the IDV size distribution? For this purpose we have sampled the R-band light curve from 2016 to 2023 which is dense enough for an IDV scale study, and has only relatively small gaps during the summer periods when OJ~287 is not visible from the ground. The full width at half-maximum of our 2021 November 12 flare is 0.5 days. Therefore we chose to find and study all flares which satisfy the condition that their full width at half-maximum is not greater than one day, which is one way to define an IDV event. 

The number distribution for the sample is shown in Figure 5. The November 12, 2021, flare is not included, as we want to consider it separately. There is a practical lower limit as to how small a variation is defined a flare. This causes a decline at the lower end of the distribution.

\begin{figure}
    \centering
    \includegraphics[angle = 0,width=0.49\textwidth]{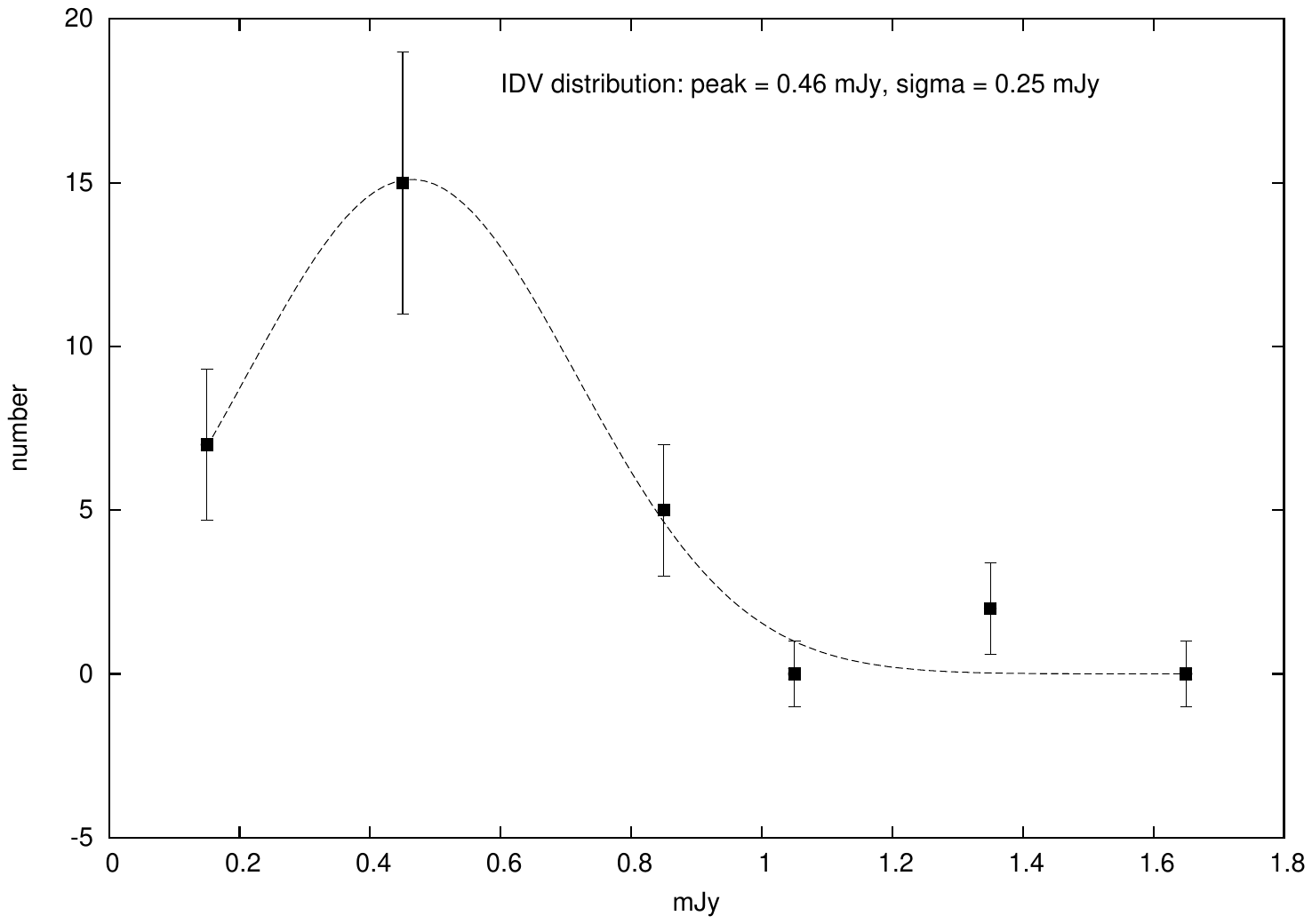}
    \caption{The number of flares observed in different size categories as a function of the mean flux density of the category. A Gaussian is fitted through the points. The 4.3 mJy flare is not shown.
}
    \label{fig:magVsPDPA1}
\end{figure}

We fit a Gaussian to the distribution as shown in Figure \ref{fig:magVsPDPA1}. We see that our flare would be about 9 standard deviations from the center of this distribution. The Gaussian tail may not be a good representation of the fall-off of the numbers with size. If we use a power-law, the numbers are expected to fall more slowly, but still we find that the 4.3 mJy flare is at least 3 standard deviations beyond what is expected. Thus it is rather unlikely that our flare is just part of the normal variation in OJ~287. As long as we regard the normal variation as a property of the primary jet, the November 21, 2021, flare with its 4.3 mJy flux does not fit in the single jet scenario.

\section{The primary and the secondary jet}
\noindent
Radiation originating in the  primary jet is seen from radio frequencies to X-rays, and the jet reaches into the megaparsec range \citep{2011ApJ...729...26M}. The jet is seen to wobble in a manner that can be explained by the influence of the secondary BH on the inner disk of the primary \citep{2012MNRAS.421.1861V,2013A&A...557A..28V,2021MNRAS.503.4400D}. The~jet is thought to point almost directly towards us, which means that occasionally the jet passes through our line of sight. It causes in a big jump in the projected direction of the jet in the sky \citep{2012ApJ...747...63A,2021MNRAS.503.4400D}. These jumps take place at different times at different frequencies, which is understandable, if the jet is helical and radiation at different frequencies arises at different parts of the helix \citep{2012MNRAS.421.1861V}.

It has been suggested several times that the secondary BH also has a jet of its own \citep{1998MNRAS.293L..13V,2013MNRAS.434.3122P,2013ApJ...764....5P,2022MNRAS.513.3165K}. In the binary BH model, the spin of the primary is rather slow \citep{2016ApJ...819L..37V}, and~consequently the jet is somewhat weak in relation to its mass. However, we have good reasons to expect that the secondary black hole spins near its maximum speed, and therefore its jet is quite~bright. This high spin would be expected because the secondary receives and accretes a new dose of gas at every disk crossing, always with the same direction of the specific angular momentum.

Following \cite{2013ApJ...764....5P}, we may estimate the number of orbital revolutions required to build up the secondary spin up to its maximum value. According to simulations by \cite{2011ApJ...731L...9I}, the number of close binary orbits leading to the present configuration of OJ~287 is of the order of one million. During each orbit the secondary BH gains about $100 M_{\odot}$ of rest mass. This is higher than the present rate since in the standard magnetic accretion disk model, the accretion rate increases almost linearly with the impact distance and since the orbit was typically an order of magnitude wider than it is today \citep{1984ApJ...277..312S,2011ApJ...731L...9I}. Therefore in a million orbits the rest mass of the secondary roughly doubles itself. This is enough to drive the spin value close to its maximum, even if started from zero \citep{1970Natur.226...64B,1974ApJ...191..507T}.

\cite{1997MNRAS.292..887G} calculate the jet luminosity in the \textit{Blandford-Znajek} process \citep{1977MNRAS.179..433B} as follows:
\begin{equation}
L_{j} \sim m^{1.1} \dot{m}^{0.8} J^2                 
\end{equation}
where $m$ is the mass of the black hole, $\dot{m}$ the mass accretion rate and $J$ is its normalized spin. For the primary, $J = 0.38$ \citep{2018ApJ...866...11D}, and for the secondary, $J \sim 1$. In the normalization, we divide by $m$, which necessarily takes a $J$ close to 1 or the maximum value, for~small $m$.
Using $\dot{m} = 1$ for the secondary (the Eddington rate) and $\dot{m} = 0.08$ for the primary \citep{2019ApJ...882...88V} makes the secondary jet $43\%$ of the total~luminosity.

During the disk-crossing the secondary has the possibility of accreting large amounts of gas from the primary disk at a rapid rate, leading to a super-Eddington rate and a large increase in brightness. On~the other hand, the~impact on the disk tends to strip the secondary disk of its outer layers, and~the disk-crossing related events should be~short-lived.

\section{Why have we not seen flares like this before?}
\noindent
We searched the optical light curve of OJ~287 during the OJ94 campaign~1993--1998,~\citep{2021yCat..41460141P}, in~the 2005--2010 campaign \citep{2011AcPol..51f..76V} and in the recent campaign (2015--2023, led by S.Z., see the Appendix), where very dense monitoring was carried out over the total period corresponding to approximately 10 years worth of data. We found that there is only one flare that is even remotely similar to the 2021.86 flare in terms of amplitude (3.3 mJy in R) and rapidity (1~day): the~1993 December flare \citep{1995A&AS..113..431K}.

According to \cite{2013ApJ...764....5P}, a huge IDV flare should arise when the secondary black hole meets dense clouds of gas during its 12-year orbit. This happens most obviously at the disk crossing. Already before the crossing, the Roche lobe of the secondary is filled with dense gas. Some part of this gas falls inside the zone of 10 Schwarzschild radii from the secondary in 0.08 yr timescale, and from there it is accreted to the secondary in about 0.02 yr. Thus we expect to see the flare about 0.1 yr after the filling of the Roche lobe. The line in Fig.\ref{fig:magVsPDPA} is copied from \cite{2013ApJ...764....5P}, using the assumption that the mass influx is directly proportional to brightness. No detailed model for the sharp flare is available from simulations. 

In order to see where this places the OJ~287 flare in 2021, we have to look at the orbit dynamics at that time. Table 2 gives orbital phases as well as observed magnitudes during certain interesting times. It is an extract from a larger orbit-linked historical light curve of OJ~287 which is found in the Appendix. The linking to the orbit is also done in the upper scale of Fig.\ref{fig:magVsPDPA} which tells the position of the secondary above the disk.

From the Appendix we may read that the secondary is 1150 AU above the midplane of the disk at 2021.756. The significance of this level above the midplane is that it has the value of the Roche lobe radius of the secondary \citep{1983ApJ...268..368E}, i.e., just at this time the secondary BH has started to swallow the gas of the primary disk. The sharp flare is observed at 2021.863. This is $\sim 0.1$ yr later, as was estimated by \cite{2013ApJ...764....5P}. As they pointed out, the final accretion spike arises in only a few orbital times. If we take the period of the innermost stable orbit (ISCO), estimated to be 3.8 hours \citep{2013MNRAS.434.3122P}, as the reference number, we find a good agreement with the 12 hour observed spike. 

Note that the time of the Roche-lobe flare is theoretically specified within an interval of $\pm0.01$ yr. The~accidental chance of detecting an exceptional flare at this time is less than $0.2\%$, since the probability of finding such a flare in the 10 yrs of monitoring data is less than~unity.

As Figure 3 illustrates, these events are only visible to us when the secondary hits the disk from above, as seen from our direction. Thus there is an opportunity to see a flare like this only once in 12 years. And to meet the opportunity, an observer has to take measurements not only on a correct night but also in the correct few-hour interval during that night. We would have missed this flare also this time, if we did not have a warning that something interesting might happen in this time frame which alerted observers to do their best for a fast sampling of the light curve.  

In order to calculate the predicted flare time at other instances, we may simply scale the Roche-lobe filling level above the disk with the distance from the primary black hole. This is because the radius of the Roche lobe scales with this distance. Also the relevant astrophysical speeds scale with the distance from the center of the accretion disk, basically due to Kepler's second law. Thus we may take the level of the secondary above the disk at the spike (see the upper scale in Figure 2), and use it as a reference number which is then scaled linearly to other impact distances. The timing is then calculated using the Appendix. The results are shown in Table 2. 

We find that in the past there has only been one observation on the night in question. This happened in 1947 when a photographic plate was taken at Sonneberg Observatory in Germany, including the position of OJ~287 in the sky, at the right time on JD2432287, obviously for other purposes. One of us (R.H.) has confirmed that the detection limit on this plate was rather poor, and one cannot see anything at the position of OJ~287. Thus the detection would not have been possible.

\begin{deluxetable}{ccccc}
\tablenum{2}
\tablecaption{Dates of Roche-lobe flares and impacts on the disk by the OJ~287 secondary BH}
\tablewidth{0pt}
\tablehead{
\colhead{Date} & \colhead{X} & \colhead{Y} & \colhead{V-mag} &\colhead{Event} 
}
\decimalcolnumbers
\startdata
1895.153 & -16007.95 & 1144.43 & - & 1895-flare\\
1895.444 & -15361.20 & 147.85 & - &  1895-impact\\ 
1902.979 & -17763.02 & 1500.39 & - & 1902-flare\\
1903.047 & -17863.07 & 1309.48 & 14.90 & observation\\
1903.390 & -18280.91 & 333.60 & - & 1903-impact\\ 
1912.442 & -10862.75 & 680.36 & - & 1912-flare\\
1912.592 & -11479.56 & 12.07 & - & 1912-impact\\
1923.627 & -6306.61 & 370.52 & - & 1923-flare\\
1923.674 & -6560.60 & 25.18 & - & 1923-impact\\
1935.372 & -4313.45 & 241.91 & - & 1935-flare\\
1935.395 & -4422.21 & 5.50 & - & 1935-impact\\
1947.225 & -3336.67 & 750.03 & 12.95 & observation\\
1947.270 & -3494.69 & 208.14 & $\geq$ 13.70 & 1947-flare\\
1947.273 & -3504.85 & 164.31 & $\geq$ 15.70 & observation\\
1947.285 & -3536.00 & 18.59 & 14.00 & 1947-impact\\
1959.196 & -3317.18 & 277.10 & 13.33 & observation\\
1959.201 & -3319.03 & 208.64 & - & 1959-flare\\
1959.219 & -3317.34 & -19.38 & - & 1959-impact\\
1971.109 & -3683.71 & 266.20 & - & 1971-flare\\
1971.129 & -3628.91 & 18.16 & 13.07 & 1971-impact\\
1982.930 & -4789.16 & 300.09 & - & 1982-flare\\
1982.960 & -4653.19 & 3.87 & 14.363 & 1982-impact\\
1994.466 & -7432.68 & 458.74 & - & 1994-flare\\
1994.537 & -7104.67 & -12.55 & - & 1994-impact\\
2004.949 & -13271.67 & 1101.68 & 15.40 & observation\\
2004.957 & -13251.91 & 1071.22 & - & 2004-flare\\
2004.963 & -13236.68 & 1047.92 & 15.19 & observation\\
2005.166 & -12676.68 & 250.00 & 14.40 & 2005-impact\\
2013.179 & -17495.82 & 1357.36 & 14.82 & observation\\
2013.183 & -17497.73 & 1346.63 & - & 2013-flare\\
2013.188 & -17500.13 & 1333.22 & 14.91 & observation\\
2013.574 & -17608.02 & 254.00 & - & 2013-impact\\
2021.859 & -11805.82 & 704.78 & 15.21 & observation\\
2021.863 & -11820.50 & 688.02 & 14.345 & 2021-flare\\
2021.872 & -11850.30 & 654.00 & 15.29 & observation\\
2022.054 & -12423.44 & -47.73 & 14.85 & 2022-impact\\
2032.614 & -6.610.41 & 402.34 & - & 2032-flare\\
2032.675 & -6.905.54 & 2.61 & - & 2032-impact\\
2044.162 & -4272.66 & 282.53 & -& 2044-flare\\
2044.192 & -4411.38 & -3.67 & - & 2044-impact\\
\enddata
\label{tab:short}
\tablecomments{Column 1: date, Columns 2 and 3: X- and Y- coordinates of the secondary BH, column 4: V-band magnitude, Column 5: Type of event. In each  impact-related group, the date of the expected Roche-lobe flare and its magnitude are given, in case observations were recorded at that night. The adjacent dates refer to nearest observations, if any. The last line is for the date of impact of the secondary BH on the midplane of the disk.}
\end{deluxetable}

\section{CONCLUSIONS}\label{sec:discuss}
\noindent
The 2021/22 observing campaign of OJ287 was planned in anticipation of a major cosmic crash: a $1.5 \times 10^{8}$ solar mass BH crashing through the accretion disk of another, bigger BH. The epoch of this event was uncertain by some months since the position of the disk in the system was not yet known. It was calculated only later. Even though such events are thought to arise regularly in OJ~287, never before had an extensive campaign been directly aimed at this particular epoch. Previously the observations were concentrated on signals that arise from the material expelled from the disk. Since such major signals were expected in July/August 2022 when the source is unobservable from the ground, the emphasis this time was on the direct signals from the impact.

What was expected, and had been discussed already several years earlier by \cite{2013ApJ...764....5P}, was the temporary activation of the jet from the $1.5 \times 10^{8}$ solar mass BH. Since the secondary black hole is 122 times smaller than the primary in the solution by \cite{2018ApJ...866...11D}, the timescales associated with the secondary jet should be correspondingly shorter. Therefore we had to look for IDV events. Purely from energetic grounds it is unlikely that the primary jet could be operated by a sufficiently small BH that it could produce major flares in the IDV time scale \citep{2023MNRAS.525.1153V}.

The 2021 November 12 (at 2 am UT) flare was fortunately well covered by the TESS campaign \citep{2024ApJ...960...11K} in the I-band (or close to it) and our campaign in the R-band. Both observed an increase of the flux by more than a magnitude in a few days, and the rise of the last half of the flare took place in only a quarter of a day. The observations allow us to compare the R-I spectral index in and out of the flare. The difference is clear: the spectrum is flatter in the flare with respect to the background by $\Delta\alpha \sim 1.0\pm0.1$. If we consider the likely separation of the base and the flare components, the difference is even clearer, $\Delta\alpha = 1.5\pm0.1$.

The large change of the spectral index around the R-band frequency is difficult to understand purely as aging of the population of relativistic electrons \citep{1970ranp.book.....P,1988AJ.....95..307I}. The situation resembles the spectral behavior normally seen at the radio frequencies \citep{1985ApJ...298..114M}. The flat-spectrum turnover frequency would then be $\sim10^{3}$ times higher than what is observed typically in quasars \citep{1992A&A...254...71V}. This could result from a different magnetic flux density and/or size of the emitting region in the secondary jet than in the primary jet.

The secondary jet origin of the radiation at this time is also deduced from the behavior of polarization, even though the coverage is not as complete as the spectral index coverage. The variability time scale puts strict limits on the size of the emission region in the IDV flare, and places it most likely inside the jet of the smaller BH.

The short life of the IDV flare, only 12 hours above the half-maximum value, makes it very difficult to detect by a single telescope on the ground. It may happen during the daytime or a cloudy period at that telescope site. Or it may just happen that there is a break in the observing schedule for some reason. Thus the Skynet Robotic Telescope Network, with telescopes on four continents, as well as the dedicated TESS satellite monitoring, were crucial to this discovery.

There are previous examples in the monitoring of OJ287 flares where the peak activity was missed for various technical reasons (e.g. the peak of the 1995 major flare and the polarization measurement of the peak of the 2005 major flare). Therefore we were fortunate this time even if we knew reasonably well what to look for and when.

The detection of the November 12, 2021, flare may also be viewed as an additional confirmation of the \cite{2018ApJ...866...11D} orbit solution. Since we are lacking a detailed theory of the secondary BH jet activation process, we could not predict to-a-day when the IDV flare should have happened. However, now that we have seen one such case, we can reasonably calculate when such events should have happened in the past. It seems that we have missed them all. Also we can give fair estimates when they will happen again: in August 2033 and in March 2044. The former event requires space based observations, but fortunately there is plenty of time to prepare for them.

\section*{ACKNOWLEDGMENTS}
\noindent
This study was based in part on observations conducted using the Perkins Telescope Observatory (PTO) in Arizona, USA, which is owned and operated by Boston University. The research at Boston University was supported in part by National Science Foundation grant AST-2108622, and a number of NASA Fermi Guest Investigator grants; the latest is 80NSSC23K1507. This work was partly funded by NCN grant No. 2018/29/B/ST9/01793 (SZ) and JSPS KAKENHI grant No. 19K03930 (KM). SC acknowledges support by ASI through contract ASI-INFN 2021-43-HH.0 for SSDC, and Instituto Nazionale di Fisica Nucleare (INFN). RH acknowledges the EU project H2020 AHEAD2020, grant agreement 871158, and internal CTU grant SGS21/120/OHK3/2T/13. ACG is partially supported by Chinese Academy of Sciences (CAS) President's International Fellowship Initiative (PIFI) (grant no. 2016VMB073). MFG is supported by the National Science Foundation of China (grant 11873073), Shanghai Pilot Program for Basic Research-Chinese Academy of Science, Shanghai Branch (JCYJ-SHFY-2021-013), the National SKA Program of China (Grant No. 2022SKA0120102), the science research grants from the China Manned Space Project with No. CMSCSST-2021-A06, and the Original Innovation Program of the Chinese Academy of Sciences (E085021002).  ZZ is thankful for support from the National Natural Science Foundation of China (grant no. 12233005). The work is partly based on observations made with the Nordic Optical Telescope, owned in collaboration by the University of Turku and Aarhus University, and operated jointly by Aarhus University, the University of Turku and the University of Oslo, representing Denmark, Finland and Norway, the University of Iceland and Stockholm University at the Observatorio del Roque de los Muchachos, La Palma, Spain, of the Instituto de Astrofisica de Canarias. This study is based in part on observations obtained with telescopes of the University Observatory Jena, which is operated by the Astrophysical Institute of the Friedrich-Schiller University. MJV acknowledges a grant from the Finnish Society for Sciences and Letters.

\bibliography{ref}{}


\bibliographystyle{aasjournal}

\end{document}